\documentstyle[aps,graphicx,amssymb,amsmath]{revtex}

\newcommand{\grad}{{\rm grad}}

\newcommand{\divz}{{\rm div}}
\newcommand{\expo}{{\rm e}}

\begin{document}

\draft

\title{The Adjoint Problem in the Presence of a Deformed Surface:
the Example of the Rosensweig Instability on Magnetic Fluids
}
\author{Adrian Lange}
\address{Universit\"at Magdeburg, Institut f\"ur Theoretische
Physik, Universit\"atsplatz 2, D-39106 Magdeburg, Germany}

\date{\today}

\maketitle

\begin{abstract}
The Rosensweig instability is the phenomenon that above a certain threshold of
a vertical magnetic field peaks appear on the free surface of a horizontal layer of
magnetic fluid. In contrast to almost all classical hydrodynamical systems, the
nonlinearities of the Rosensweig instability are entirely triggered by the properties
of a deformed and a priori unknown surface. The resulting problems in defining an
adjoint operator for such nonlinearities are illustrated. The implications concerning
amplitude equations for pattern forming systems with a deformed surface are discussed. 
\end{abstract}

\section{Introduction}
\label{sec:1}

Magnetic fluids (MFs) are stable colloidal suspensions of ferromagnetic nanoparticles
dispersed in a non-magnetic carrier liquid. The nanoparticles are coated with a
layer of chemically adsorbed surfactants to avoid agglomeration. The
behaviour of MFs is characterized by the intricate interaction of their
hydrodynamic and magnetic properties with external forces. This complex
interaction causes  many fascinating phenomena, as the `negative
viscosity' effect and the Weissenberg effect (for a review see \cite{odenbach00_viscous})
or as the labyrinthine instability and the Rosensweig instability
\cite{rosensweig}. The latter instability occurs when a horizontal
layer of MF with a free surface is subjected to a uniform and vertically
oriented magnetic field. Above a certain threshold of the magnetic field
that surface becomes unstable, giving rise to a hexagonal pattern of peaks
\cite{rosensweig,cowley67}. Despite the fact that the Rosensweig instability has
been known for many decades, some aspects have been addressed only recently:
the inclusion of the fluid viscosity \cite{salin93} and of a finite
layer thickness \cite{weilepp96}, the hexagon-square transition
\cite{abou00} or the wave number selection problem
\cite{abou00,lange00_wave}.

The {\it quantitative} comparison of theoretical and experimental results is
presently limited to linear aspects of the Rosensweig instability.
Convincing quantitative agreement is found for the wave number
of maximal growth \cite{lange00_wave}, for the parametric stabilization of the
Rosensweig instability \cite{petrelis00}, and for the wave resistance in magnetic
fluids \cite{browaeys01}. For the nonlinear aspects, comparisons are
restricted to the detection of the same {\it qualitative} features. Stability
regions for hexagons and squares are found in the theory
\cite{kuznetsov76,gailitis77,beyer80,twombly80,twombly83,silber88,friedrichs01}
and in the experiment \cite{abou00}. The experimentally observed increase in the
wavelength of the emerging hexagonal pattern \cite{abou00} is detectable
as well in a theoretical analysis \cite{friedrichs01}. The main reason for
lacking {\it quantitative} comparisons is the difference between the susceptibility
of the experimentally used fluids and the susceptibility up to which the nonlinear
analyses are valid. Finite amplitudes for all three patterns of hexagons, squares, and rolls
are given for susceptibilities $\chi$ smaller then $0.41$ for an infinitely thick
layer (see Figs.~3 and 6 in \cite{friedrichs01}). But the experiments were performed
with a magnetic fluid of $\chi_{{\rm exp}}=1.4$ \cite{abou00}. Pattern selection studies
with fluids of different susceptibilities have not yet been carried out.

Weakly nonlinear analyses of the Rosensweig instability were done by means of an
energy minimization principle \cite{kuznetsov76,gailitis77,friedrichs01}, by
methods of functional analysis \cite{twombly80,twombly83,silber88}, by a
generalized Swift-Hohenberg equation \cite{herrero94,kubstrup96}, and
by a multiple-scale analysis \cite{malik84,malik93}. The first two approaches
are suited only for static problems and the Swift-Hohenberg equation lacks
coefficients containing the fluid properties and the geometry of the system.
Consequently, amplitude equations stemming from a multiple-scale analysis 
based on the fundamental hydrodynamic equations are customized for the study of static
{\it and} dynamical problems as well as for the quantitative comparison with the
experiment. The standard route of the multiple-scale analysis is modified in
\cite{malik84,malik93} to circumvent Fredholms theorem, i.e. the definition
of an adjoint operator. In the present paper a multiple-scale analysis is presented
which involves the expansion of all physical quantities at the deformed surface.
The resulting problem in defining an adjoint operator and the
subsequent consequences are the main purpose of this paper. It is organized as
follows: the system and the relevant equations of the problem are displayed in the
next section. Based on the governing equations and the boundary conditions the
different character of the nonlinearities of the Rosensweig instability is
emphasised (Sec.~\ref{sec:2.3}). Whereas the linear problem is shortly recapitulated
in Sec.~\ref{sec:3}, the adjoint problem is addressed in detail in Sec.~\ref{sec:4}.
In the final section~\ref{sec:5}, the problems concerning amplitude equations
for pattern forming systems with a deformed surface are discussed as well as open
questions and further prospects are outlined.

\section{System and Equations of the Problem}
\label{sec:2}

\begin{figure}[htbp]
  \begin{center}
    \includegraphics[scale=0.45]{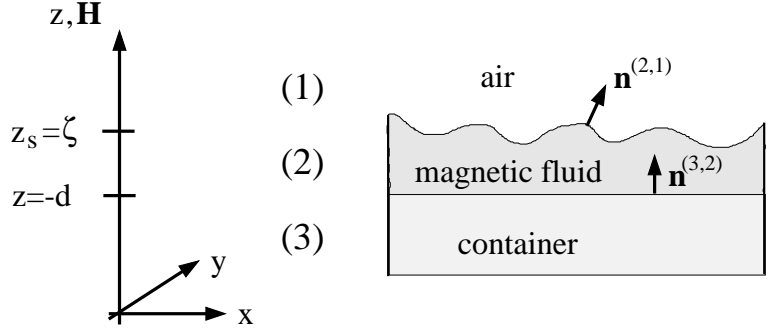}
    \caption{Sketch of the system.}
    \label{fig:1}
  \end{center}
\end{figure}

A horizontally unbounded layer of an incompressible, nonconducting, and viscous
magnetic fluid of finite thickness $d$ and constant density $\rho$ is considered.
The MF is bounded from below ($z=-d$) by the bottom of a container
made of a magnetically impermeable material and has a free surface described by
$z_s=\zeta (x, y, t)$ with air above. The electrically insulating
fluid justifies the stationary form of the Maxwell equations, which reduce
to the Laplace equation for the magnetic potentials $\Phi^{(i)}$ in each of
the three different regions. Upper indices denote the considered media:
$(1)$ air, $(2)$ magnetic fluid, and $(3)$ container (Fig.~\ref{fig:1}).
It is assumed that the magnetization ${\bf M}^{(2)}$ of the magnetic fluid depends
linearly on the magnetic field ${\bf H}^{(2)}$, ${\bf M}^{(2)} =(\mu_r -1){\bf
H}^{(2)}$, where $\mu_r$ is the relative permeability of the fluid.
Additionally, the magnetization is considered to be a linear function of
the density $\rho^{(2)}$ which results in the usual form of the Kelvin force
${\bf F}_{{\rm K}}=\mu_0\bigr( {\bf M}^{(2)}\grad\bigr) {\bf H}^{(2)}$ \cite{rosensweig}.
The system is governed by the equation of continuity, the
Navier-Stokes equations for the magnetic fluid,
\begin{eqnarray}
  \label{eq:1}
  \divz\, {\bf v}^{(2)} & = & 0 \;,\\
  \label{eq:2}
  \rho^{(2)}\partial_t {\bf v}^{(2)} + \rho^{(2)}\bigr( {\bf v}^{(2)}\grad\bigr){\bf v}^{(2)}
  &=&-\grad\, p^{(2)} +\mu^{(2)}\Delta {\bf v}^{(2)}
  +\mu_0\bigr( {\bf M}^{(2)}\grad\bigr) {\bf H}^{(2)} + \rho^{(2)}{\bf g}\; ,
\end{eqnarray}
and the Laplace equation in each medium,
\begin{equation}
  \label{eq:3}
  \Delta\Phi ^{(i)} = 0\; .
\end{equation}
The scalar magnetic potentials are defined by ${\bf H}^{(i)}=-\grad \Phi^{(i)}$. The velocity
field in the MF is denoted by ${\bf v}^{(2)}$, the dynamic viscosity by $\mu^{(2)}$, the
pressure field by $p^{(2)}$, and the acceleration due to gravity by ${\bf g}=(0,0,-g)$. The
first three terms on the right-hand side of Eq.~(\ref{eq:2}) result from
${\rm div}\,\overleftrightarrow{T}^{(2)}$, where the components of the stress tensor
$\overleftrightarrow{T}^{(2)}$ read \cite{rosensweig}
\begin{equation}
  \label{eq:4}
  T_{ij}^{(2)} = \left\{ -p^{(2)}
   - \mu_0{\bigr( H^{(2)}\bigr) ^2\over 2}\right\}\delta_{ij}
   + H_i^{(2)} B_j^{(2)}  +\mu^{(2)}\left(\partial_i v_j^{(2)} +\partial_j v_i^{(2)} \right)\, .
\end{equation}
$M^{(2)}$, $H^{(2)}$, and $B^{(2)}$ denote the absolute value of the magnetization,
the magnetic field, and the induction ${\bf B}^{(2)}$ in the MF. Since the final
arrangement of peaks is a static configuration, the velocity field in the MF is
set to zero, ${\bf v}^{(2)}=0$, and the surface depends only on the horizontal spatial
coordinates, $z_s =\zeta(x,y)$. Applying all assumptions, the static form of
the Navier-Stokes equation is
\begin{equation}
  \label{eq:5}
  0 = -\grad\, p^{(2)} + {\mu_0\over 2}(\mu_r-1)\,\grad \bigr( H^{(2)}\bigr)^2
      - \rho^{(2)}g{\bf e}_z\; ,
\end{equation}
where ${\bf e}_z=(0,0,1)$ is the unit vector in $z$-direction. The governing equations have to be
supplemented by the appropriate boundary conditions. These are the continuity of the normal
(tangential) component of the magnetic induction (magnetic field) at the top and bottom interface,
\begin{eqnarray}
  \label{eq:7}
  &&{\bf n}^{(2,1)}\cdot\left( {\bf B}^{(1)}-{\bf B}^{(2)}\right) =0,\quad
  {\bf n}^{(2,1)}\times\left( {\bf H}^{(1)}-{\bf H}^{(2)}\right) =0
  \quad {\rm at~} z=z_s\; ,\\
  \label{eq:8}
  &&{\bf n}^{(3,2)}\cdot\left( {\bf B}^{(2)}-{\bf B}^{(3)}\right) =0,\quad
  {\bf n}^{(3,2)}\times\left( {\bf H}^{(2)}-{\bf H}^{(3)}\right) =0
  \quad {\rm at~} z= -d \; ,
\end{eqnarray}
and the continuity of the normal component of the stress tensor across the
free surface
\begin{equation}
  \label{eq:9}
  \left( \overleftrightarrow{T}^{(1)} - \overleftrightarrow{T}^{(2)}\right){\bf n}^{(2,1)}
  -\sigma K {\bf n}^{(2,1)} = 0\qquad\qquad  {\rm at~} z=z_s\; .
\end{equation}
In (\ref{eq:9}) the surface tension between the magnetic fluid and air is denoted by
$\sigma$, the curvature of the surface by $K={\rm div}\, {\bf n}^{(2,1)}$, and the
unit vector normal to the MF surface by
\begin{equation}
  \label{eq:10}
  {\bf n}^{(2,1)} = {{\rm grad}\left[ z-\zeta (x,y)\right]\over |{\rm grad}\left[ z-
  \zeta (x,y)\right]|} ={(-\partial_x\zeta , -\partial_y\zeta , 1)\over
  \sqrt{1+(\partial_x\zeta)^2 + (\partial_y\zeta)^2}}\, .
\end{equation}
The upper index $(2,1)$ at the unit vector indicates that ${\bf n}^{(2,1)}$ points from
medium 2 towards medium 1; analogous for the normal vector ${\bf n}^{(3,2)}=(0,0,1)$
(see Fig.~\ref{fig:1}). The difference of the tangential components of the stress tensor is
identically zero because of the continuity of the magnetic fields and inductions in
(\ref{eq:7}). Neglecting the influence of the air pressure with respect to the fluid
pressure, $p^{(1)}\simeq 0$, and using ${\bf M}^{(1)}\equiv 0$, one finally gets from
Eq.~(\ref{eq:9})
\begin{equation}
  \label{eq:11}
  {\mu_0\over 2}\left[ {\bf M}^{(2)}{\bf n}^{(2,1)}\right]^2
  + p^{(2)} - \sigma K = 0\qquad\qquad  {\rm at~} z= z_s\; ,
\end{equation}
where $p^{(2)}$ is the solution of Eq.~(\ref{eq:5}).

\subsection{Basic state}
\label{sec:2.1}
As long as the applied spatially homogeneous magnetic field perpendicular
to the surface,
\begin{equation}
  \label{eq:12}
  {\bf H}^{(i)}={\bf H}_G^{(i)}=(0,0,H_G^{(i)})\qquad
  {\bf B}^{(i)}={\bf B}_G^{(i)}=(0,0,B_G^{(i)})\qquad
  {\bf M}^{(2)}={\bf M}_G^{(2)}=(0,0,M_G^{(2)})\; ,
\end{equation}
is below a certain strength, the system is in its basic or ground state. This state
is given by the plane surface $z_s = 0$. The corresponding solution for the
fluid pressure is
\begin{equation}
  \label{eq:13}
  p_G^{(2)}=-\rho^{(2)} g z - {\mu_0\over 2} \bigr( M_G^{(2)}\bigr)^2\; ,
\end{equation}
where the constant resulting from (\ref{eq:5}) was determined by (\ref{eq:11}) and
$H_G^{(2)}=H_G^{(2)}\bigr|_0$ was used. Here $\bigr|_0$ denotes the evaluation at the plane
interface $z_s =0$. The boundary conditions (\ref{eq:7}, \ref{eq:8}) are fulfilled by
$B_G^{(1)}=B_G^{(2)}$ and $B_G^{(2)}=B_G^{(3)}$, respectively.

\subsection{Small disturbances and their expansion}
\label{sec:2.2}
In order to study the stability of the basic state and the pattern selection problem
in the weakly nonlinear regime, small deviations from the basic state are considered
\begin{eqnarray}
  \nonumber
  z_s&=&0+\zeta \;,\hskip 1.97cm
  {\bf H}^{(i)}={\bf H}_G^{(i)}+{\bf h}^{(i)}\; ,\hskip 0.65 cm
  {\bf B}^{(i)}={\bf B}_G^{(i)}+{\bf b}^{(i)}\; ,\\
  \label{eq:14}
  {\bf M}^{(2)}&=&{\bf M}_G^{(2)}+{\bf m}^{(2)}\; ,\qquad
  \Phi^{(i)}=\Phi_G^{(i)}+\phi^{(i)}\; ,\qquad
  p^{(2)}=p_G^{(2)}+\pi^{(2)}\; .
\end{eqnarray}
Dimensionless quantities are introduced, where physical quantities are denoted by
a hat in the rest of the paper. All quantities associated with the magnetic
field are scaled by the critical magnetic induction in the limit of an infinite thickness
of the MF layer, $\hat B_{c,\infty}$. This preference is based on the fact that magnetic
inductions can be directly measured by Hall probes. Furthermore, it is
assumed that the deviations are proportional to the applied external induction. Thus
the following scaling is used
\begin{eqnarray}
  \label{eq:15}
  \hat{\bf B}_G^{(i)}&=&{\bf B}_{{\rm ext}} \hat B_{c,\infty}\; ,
  \hskip 3.45 cm
  \hat{\bf b}^{(i)}={\bf b}^{(i)}B_{{\rm ext}} \hat B_{c,\infty}\; ,\\
  \label{eq:16}
  \hat{\bf M}_G^{(i)}&=&{\bf B}_{{\rm ext}}{(\mu_r-1)\over \hat\mu_0\mu_r^{(i)}} \hat B_{c,\infty}\; ,
  \hskip 2.03 cm
  \hat{\bf m}^{(i)}={\bf m}^{(i)}B_{{\rm ext}}{(\mu_r-1)\over \hat\mu_0\mu_r^{(i)}}\hat B_{c,\infty}\; ,\\
  \label{eq:17}
  \hat{\bf H}_G^{(i)}&=&{\bf B}_{{\rm ext}}{\hat B_{c,\infty}\over \hat\mu_0\mu_r^{(i)}}\; ,
  \hskip 3.26 cm
  \hat{\bf h}^{(i)}={\bf h}^{(i)}B_{{\rm ext}}{\hat B_{c,\infty}\over \hat\mu_0\mu_r^{(i)}}\; ,\\
  \label{eq:18}
  \hat\pi^{(2)}&=&\pi^{(2)}B_{{\rm ext}}^2{2(\mu_r+1)\over \mu_r (\mu_r-1)^2}\sqrt{\hat\rho^{(2)}\hat g
  \hat\sigma}\; ,
  \hskip 1.0 cm
  \hat l = {l\over \hat k_{c,\infty}}\; .
\end{eqnarray}
For an infinite thickness of the layer, the critical induction and critical wave number, respectively,
are \cite{cowley67}
\begin{equation}
  \label{eq:19}
  \hat B_{c,\infty}^2 = {2\hat\mu_0\,\mu_r (\mu_r +1)\sqrt{\hat\rho^{(2)}\hat\sigma\,\hat g}
  \over (\mu_r -1)^2} \qquad\qquad
  \hat k_{c,\infty}=\sqrt{{\hat\rho^{(2)}\hat g\over \hat\sigma}}\; .
\end{equation}
The dimensionless quantity $B_{{\rm ext}}$ in ${\bf B}_{{\rm ext}}=(0, 0, B_{{\rm ext}})$ measures
the strength of the applied external induction in units of $\hat B_{c,\infty}$.

With this scaling the solution of the Navier-Stokes equation (\ref{eq:5}) becomes
\begin{equation}
  \label{eq:20}
  \pi^{(2)}= {\eta\over 1-\eta}\left[ 2 h_z^{(2)}+\bigr( {\bf h}^{(2)}\bigr)^2\right] + c\; ,
\end{equation}
with a yet unknown constant $c$. The Laplace equation~(\ref{eq:3}) in each medium is
\begin{equation}
  \label{eq:21}
  \Delta \phi^{(i)}=0 \; ,
\end{equation}
and the boundary conditions~(\ref{eq:7}, \ref{eq:8}, \ref{eq:11}) are
\begin{eqnarray}
  \label{eq:22}
  {\rm at~}z=-d\qquad
  0 &=& -\phi^{(2)} + \phi^{(3)}\; ,\\
  \label{eq:23}
  {\rm at~}z=-d\qquad
  0 &=& -{1+\eta\over 1-\eta}\partial_z\phi^{(2)} + \partial_z\phi^{(3)}\; ,\\
  \label{eq:24}
  {\rm at~}z=\zeta\qquad
  0 &=& -{2\eta\over 1-\eta}\zeta + \phi^{(1)} - \phi^{(2)}\; ,\\
  \nonumber
  {\rm at~}z=\zeta\qquad
  0 &=& -\partial_x \zeta \left( -\partial_x \phi^{(1)}+{1+\eta\over 1-\eta}\,\partial_x \phi^{(2)}\right)
  -\partial_y \zeta \biggr( -\partial_y \phi^{(1)}+{1+\eta\over 1-\eta}\,\partial_y \phi^{(2)}\biggr)\\
  \label{eq:25}
  && -\partial_z \phi^{(1)}+{1+\eta\over 1-\eta}\,\partial_z \phi^{(2)}\; ,\\
  \nonumber
  {\rm at~}z=\zeta\qquad
  0 &=&-{2 B_{{\rm ext}}^2\over (1+\eta)}{\bigr[ \left(\partial_x \zeta\right)^2+\left(\partial_y \zeta\right)^2\bigr]
  \over \bigr[ 1+\left(\partial_x \zeta\right)^2+\left(\partial_y \zeta\right)^2\bigr]}\\
  \nonumber
  && +{2 B_{{\rm ext}}^2 (1-\eta)\over \eta(1+\eta)}\left( {2\eta\over (1-\eta)}{{\bf h}^{(2)}{\bf n}^{(2,1)}\over
  \sqrt{ 1+\left(\partial_x \zeta\right)^2+\left(\partial_y \zeta\right)^2}}+ {\bf h}^{(2)}{\bf e}_z\right)\\
  \label{eq:26}
  &&+{B_{{\rm ext}}^2 (1-\eta)\over \eta(1+\eta)}\left[ {2\eta\bigr({\bf h}^{(2)}{\bf n}^{(2,1)}
  \bigr)^2\over (1-\eta)}+\bigr({\bf h}^{(2)}\bigr)^2\right] - z - K +{B_{{\rm ext}}^2 (1-\eta)^2\over
  \eta^2(1+\eta)}\,c \;.
\end{eqnarray}
The curvature of the surface is
\begin{equation}
  \label{eq:27}
  K={-\partial_{xx}\, \zeta -\partial_{yy}\, \zeta \over
    \sqrt{1+(\partial_x \zeta)^2+(\partial_y \zeta)^2}}+
    {(\partial_x\zeta )^2 \partial_{xx}\, \zeta +2 \partial_x\zeta \,\partial_y \zeta\,
    \partial_{xy} \zeta + (\partial_y\zeta )^2\,\partial_{yy}\, \zeta
    \over \bigr[ 1+(\partial_x \zeta)^2+(\partial_y \zeta)^2 \bigr]^{3/2}}\; ,
\end{equation}
and the widely used quantity
\begin{equation}
  \label{eq:28}
  \eta ={\mu_r-1\over \mu_r +1}
\end{equation}
was introduced $\bigr($$\gamma=(3/4)\eta$ in \cite{gailitis77}, \cite{friedrichs01,engel99_non}$\bigr)$.

Since the disturbances of the magnetic field are located in the vicinity of
the magnetic fluid, they have to disappear as $z$ goes to infinity, i.e. the
magnetic field disturbances have to fulfill two more conditions:
\begin{eqnarray}
\label{eq:29}
  {\rm at~} z= -\infty\qquad 0 &=& \phi^{(3)}\qquad \; ,\\
\label{eq:30}
  {\rm at~} z= \infty\hskip 0.97 cm 0 &=& \phi^{(1)}\qquad \; .
\end{eqnarray}
In summary, the system is governed by three differential equations of second order (\ref{eq:21})
and the corresponding six boundary conditions (\ref{eq:22}--\ref{eq:25}, \ref{eq:29}, \ref{eq:30})
for the magnetic potentials. The remaining equation (\ref{eq:26}) is a differential
equation for $\zeta(x,y)$ whose solution determines the shape of the surface.

To derive solutions with a finite but small amplitude, the Eqs.~(\ref{eq:21}-\ref{eq:26}) are solved
perturbatively. The external control parameter $B_{{\rm ext}}$ and the various physical quantities are
expanded
\begin{eqnarray}
  \label{eq:31}
  B_{{\rm ext}}^2 &=& B_c^2 + \varepsilon\,B_1+\varepsilon^2\,B_2+\cdots\; ,\\
  \label{eq:32}
  X &=& \varepsilon\,X_0 + \varepsilon^2\,X_1+\varepsilon^3\,X_2+\cdots\; ,
\end{eqnarray}
where $X$ can be any of the quantities $\phi^{(i)}$, $\zeta$, and $c$. $\varepsilon$ is an expansion
parameter with respect to the magnitude of the surface deformation ($0<\varepsilon \ll 1$). The boundary
conditions (\ref{eq:24}-\ref{eq:26}) have to be evaluated at an a priori unknown surface $\zeta(x,y)$.
Therefore it is convenient to expand all potentials $\phi^{(i)}$ on the surface in terms of $\phi^{(i)}$
at $z=0$,
\begin{equation}
  \label{eq:33}
  \phi^{(i)}(x,y,z=\zeta) = \phi^{(i)}(x,y,0)+\partial_z \phi^{(i)}(x,y,z)\Bigr|_0\cdot\zeta
     +{1\over 2} \partial_{zz}\phi^{(i)}(x,y,z)\Bigr|_0\cdot\zeta^2 + \cdots\; ,
\end{equation}
and similarly for all space derivatives of $\phi^{(i)}$. The chosen expansion of the external
parameter follows the way as it is known from thermal convection problems of ordinary fluids
\cite{busse82,godreche98} or magnetic fluids \cite{zebib96} and from the B\' enard-Marangoni
problem \cite{golovin97,bragard98,engel00_mara}. The expansion (\ref{eq:31}, \ref{eq:32})
describes a situation slightly above the threshold of the linear instability and differs from
those used in \cite{malik84,malik93}, where the magnetic field $H$, acting as external parameter, is not
expanded. The consequences of this difference will be discussed in Sec.~\ref{sec:5}.

\subsection{Different nonlinearity}
\label{sec:2.3}
Before going into the details of the expansion procedure, it is illuminating to analyze the set of
Eqs.~(\ref{eq:21}--\ref{eq:26}). The differential equations (\ref{eq:21}) and the boundary conditions
at the bottom of the container (\ref{eq:22}, \ref{eq:23}) are {\it linear} equations. Only the three
remaining boundary equations (\ref{eq:24}--\ref{eq:26}) contain {\it nonlinear} terms.
In contrast to almost all classical hydrodynamic systems, the nonlinearity stems from
the boundary conditions and not from the differential equations which describe phenomena in
the bulk of the system. For example in convection systems  the nonlinear
contributions are caused by the advective term, i.e. by a bulk term, in the Navier-Stokes
equation and in the equation of head conduction. For the Rosensweig instability the nonlinearities
are caused by the {\it deformed} and {\it unknown} surface $\zeta$ at which the potentials
have to be calculated. Thus each linear term in $\phi^{(1)}$ and $\phi^{(2)}$ in
Eqs.~(\ref{eq:24}-\ref{eq:26}) generates nonlinear terms via Eq.~(\ref{eq:33}). A further nonlinear
contribution comes from the Kelvin force density in which a term quadratic in the potential $\phi^{(2)}$
appears, to determine again at the deformed and unknown surface. This qualitatively different character
of the Rosensweig instability makes it fascinating as well as hard to treat the instability with the
classical methods. One indicator of the essential role of the boundaries is the fact that
the external parameter $B_{{\rm ext}}$ appears only in the boundary condition~(\ref{eq:26}).

Performing the expansion according to (\ref{eq:31}--\ref{eq:33}) in (\ref{eq:21}--\ref{eq:26})
generates a hierarchy of linear equations for $U_0$, $U_1$, and $U_2$
\begin{eqnarray}
  \label{eq:41}
  L_0 U_0 &=&0\; ,\\
  \label{eq:42}
  L_0 U_1 &=& -L_1 U_0 + N_1(U_0, U_0)\; ,\\
  \label{eq:43}
  L_0 U_2 &=& -L_1 U_1 -L_2 U_0 + N(U_0,U_1) + N(U_1,U_0) + N(U_0,U_0,U_0) +N_2(U_0,U_0)\; ,
\end{eqnarray}
by matching the three lowest powers of $\varepsilon$. The vector
\begin{equation}
\label{eq:44}
U=(\phi^{(3)},  \phi^{(2)},  \phi^{(1)}, \phi^{(1)}\bigr|_0,  \phi^{(2)}\bigr|_0, \zeta)^T
=\varepsilon U_0 +\varepsilon^2 U_1 +\varepsilon^3 U_2 +\,\cdots
\end{equation}
is the augmented state vector, $L_0$, $L_1$, and $L_2$ are linear operators, and the vectors
$N(\cdot )$, $N_1(\cdot )$, and $N_2(\cdot )$ contain the nonlinear contributions. With the above mentioned,
it is clear that in each order of $\varepsilon$ different nonhomogeneous boundary problems have to be solved.
This again indicates the different quality of the problem considered here in comparison to classical
convection problems. For the boundary
conditions, terms involving $\phi^{(1)}$, $\phi^{(2)}$, and their derivatives have to be evaluated
at $z=0$  wherefore the vector $U$ contains the potentials $\phi^{(1)}$ and $\phi^{(2)}$ at $z=0$ 
additionally to the potentials itself. Similar state vector were used for the B\'enard-Marangoni
system \cite{bragard98,engel00_mara} or for the electroconvection problem in a thin
film \cite{deyirmenjian97}.

\section{Linear problem}
\label{sec:3}
The linear stability problem and the corresponding adjoint problem play a fundamental role
in the nonlinear analysis. Since the linear problem has been considered in details elsewhere
\cite{salin93,weilepp96,lange00_wave,abou97,mueller98,lange01_growth} only some of its more
important aspects are recapitulated here.

Inserting the ansatz (\ref{eq:31}-\ref{eq:33}) into (\ref{eq:21}--\ref{eq:26}) gives at
order $O(\varepsilon)$
\begin{eqnarray}
  \label{eq:45}
  0 &=& \Delta \phi_0^{(i)}\; ,\\
  \label{eq:46}
  {\rm at~}z=-d\qquad
  0 &=& -\phi_0^{(2)} + \phi_0^{(3)}\; ,\\
  \label{eq:47}
  {\rm at~}z=-d\qquad
  0 &=& -{1+\eta\over 1-\eta}\partial_z\phi_0^{(2)} + \partial_z\phi_0^{(3)}\; ,\\
  \label{eq:48}
  {\rm at~}z=0\qquad
  0 &=& -{2\eta\over 1-\eta}\zeta_0 + \phi_0^{(1)} - \phi_0^{(2)}\; ,\\
  \label{eq:49}
  {\rm at~}z=0\qquad
  0 &=& -\partial_z \phi_0^{(1)}+{1+\eta\over 1-\eta}\,\partial_z \phi_0^{(2)}\; ,\\
  \label{eq:50}
  {\rm at~}z=0\qquad
  0 &=& -{2(1-\eta) B_c^2\over \eta (1+\eta)} \partial_z \phi_0^{(1)} - \zeta_0
  +\left(\partial_{xx} +\partial_{yy}\right)\zeta_0 + {(1-\eta)^2 B_c^2\over \eta^2 (1+\eta)}c_0\; .
\end{eqnarray}
The equations (\ref{eq:45}-\ref{eq:49}) are solved by
\begin{eqnarray}
  \label{eq:51}
  \phi_0^{(3)} &=& -{\eta (1+\eta)\expo^{k(d+z)}\over \expo^{kd}-\eta^2\expo^{-kd}}
     \bar\zeta_0\expo^{i{\bf kr}}\; ,\\
  \label{eq:52}
  \phi_0^{(2)} &=& -{\eta\left[ \expo^{k(d+z)}+\eta\expo^{-k(d+z)}\right]\over \expo^{kd}-\eta^2\expo^{-kd}}
     \bar\zeta_0\expo^{i{\bf kr}}\; ,\\
  \label{eq:53}
  \phi_0^{(1)} &=& {\eta (1+\eta)\left[ \expo^{k(d-z)}-\eta\expo^{-k(d+z)}\right]\over (1-\eta)
     \left[\expo^{kd}-\eta^2\expo^{-kd}\right]} \bar\zeta_0\expo^{i{\bf kr}}\; ,\\
  \label{eq:54}
  \zeta_0 &=& \bar\zeta_0\expo^{i{\bf kr}}\; ,
\end{eqnarray}
with ${\bf k}=(k_x,k_y)$, $k=\sqrt{k_x^2+k_y^2}$, ${\bf r}=(x,y)$, and $\bar\zeta_0$ the
amplitude of the surface deformation.
Inserting this solution into (\ref{eq:50}) leads to the known dispersion relation
for an inviscid magnetic fluid of finite thickness in dimensionless quantities
\cite{weilepp96,lange00_wave,abou97,mueller98}
\begin{equation}
  \label{eq:55}
  1+k^2-2 B_c^2 k\left({\expo^{kd}-\eta\expo^{-kd}\over \expo^{kd}-\eta^2\expo^{-kd}}\right) =0\; ,
\end{equation}
providing that the unknown constant $c_0$ is set to zero. By solving equation
(\ref{eq:55}) numerically for a given thickness $d$, one gets the height dependent
critical values $B_c(d)$ and $k_c(d)$ at the onset of the instability.

The linear operator $L_0$ together with the corresponding augmented state vector $U_0$ of
Eq.~(\ref{eq:41}) are defined on the basis of Eqs.~(\ref{eq:45}, \ref{eq:48}-\ref{eq:50}):
\begin{equation}
  \label{eq:56}
  L_0 U_0 = 0 =
  \begin{pmatrix}
    \Delta & 0      & 0      & 0                 &  0  & 0 \\
    0      & \Delta & 0      & 0                 &  0  & 0 \\
    0      & 0      & \Delta & 0                 &  0  & 0 \\
    0      & 0      & 0      & 1                 & -1  & -{2\eta\over 1-\eta} \\
    0      & {1+\eta\over 1-\eta}\partial_z\bigr|_0 & -\partial_z\bigr|_0    & 0 & 0 & 0 \\
    0      & 0      & -{2(1-\eta)\over \eta (1+\eta)}B_c^2\partial_z\bigr|_0 & 0 & 0 & \partial_{xx}+\partial_{yy}-1
  \end{pmatrix}
  \begin{pmatrix}
    \phi_0^{(3)}\\
    \phi_0^{(2)}\\
    \phi_0^{(1)}\\
    \phi_0^{(1)}\bigr|_0\\
    \phi_0^{(2)}\bigr|_0\\
    \zeta_0
  \end{pmatrix}\; .
\end{equation}

\section{Adjoint problem}
\label{sec:4}

To proceed with the derivation towards an amplitude equation, one has to solve
the linear inhomogeneous equations for $U_1$ and $U_2$, Eqs.~(\ref{eq:42}, \ref{eq:43}).
According to Fredholms theorem \cite{cross93}, they have a solution if and only if the
right hand side is orthogonal to the zero space of the linear operator $L_0$. If
$\bar U_0$ is an eigenvector of the adjoint linear operator $L_0^\dagger$ with zero
eigenvalue, then multiplying Eq.~(\ref{eq:42}) from left with $\bar U_0$ reads
$\bigr($analogous for Eq.~(\ref{eq:43})$\bigr)$
\begin{equation}
\label{eq:57}
-<\bar U_0, L_1 U_0> + <\bar U_0, N_1(U_0, U_0)> = <\bar U_0, L_0 U_1> =
< L_0^\dagger\bar U_0, U_1 > = 0\; .
\end{equation}
The scalar product is denoted by $<\,,\,>$,
where the explicit form depends on the considered problem. Applying these
solvability conditions, by which $B_1$ and $B_2$ $\bigr($see Eq.~(\ref{eq:31})$\bigr)$
can be expressed, leads to  the general amplitude equation for a pattern selection problem.

In Sec.~\ref{sec:2.3} the different character is described of the pure `surface-nonlinearities' 
occuring in the Rosensweig instability. The question arises whether the established
methods for `bulk-nonlinearities', as for example for the Rayleigh-B\' enard convection,
will work here as well. Motivated by \cite{bragard98,engel00_mara,deyirmenjian97}, where
nonlinearities triggered by the bulk and the plain surface appear, the following scalar product
\begin{eqnarray}
\nonumber
  <\bar U_0, U> &=& \lim_{l\rightarrow\infty}{1\over l^2}\int_{-l/2}^{l/2}dx \int_{-l/2}^{l/2}dy
  \biggr[ \int_{-\infty}^{-d}\!\!dz\, \bar\phi_0^{(3)\ast}\phi^{(3)}
  +{1+\eta\over 1-\eta} \int_{-d}^{0}\!\!dz\, \bar\phi_0^{(2)\ast}\phi^{(2)}
  +\int_{0}^{\infty}\!\!dz\, \bar\phi_0^{(1)\ast}\phi^{(1)}\\
\label{eq:62}
  &&
  +d\bar u_{0,4}^\ast\bigr|_0\,\phi^{(1)}\bigr|_0
  +e\bar u_{0,5}^\ast\bigr|_0\,\phi^{(2)}\bigr|_0
  +f\bar u_{0,6}^\ast\bigr|_0\,\zeta\biggr]
\end{eqnarray}
is used. $\bar U_0$ is chosen since Eqs.~(\ref{eq:42}, \ref{eq:43}) have to be multiplied by
$\bar U_0$. The first three terms in the square brackets in Eq.~(\ref{eq:62}) are contributions
from the volume, whereas the last three terms are contributions from the surface. The possible
prefactors $d$, $e$, and $f$ schwach nichtlinear as well as the surface components of the adjoint vector $\bar U_0$,
$\bar u_{0,4}$, $\bar u_{0,5}$, and $\bar u_{0,6}$, have yet to be determined. From \cite{bragard98,engel00_mara}
it is known that the surface components of the adjoint state vector are not just simply the adjoint
components of the state vector. Using the identity
\begin{eqnarray}
\nonumber
  \bar\phi_0^{(i)\ast}\Delta \phi^{(i)} &=& 
   \partial_x \left[ \bar\phi_0^{(i)\ast}\, \partial_x\phi^{(i)} -\partial_x \bar\phi_0^{(i)\ast}\, \phi^{(i)}\right]
  +\partial_y \left[ \bar\phi_0^{(i)\ast}\, \partial_y\phi^{(i)} -\partial_y \bar\phi_0^{(i)\ast}\, \phi^{(i)}\right] 
  +\partial_{xx} \bar\phi_0^{(i)\ast}\, \phi^{(i)}\\
\label{eq:63}
  &&+\partial_{yy} \bar\phi_0^{(i)\ast}\, \phi^{(i)}+\bar\phi_0^{(i)\ast}\,\partial_{zz} \phi^{(i)}\; ,
\end{eqnarray}
and the conditions (\ref{eq:22}, \ref{eq:23}, \ref{eq:29}, \ref{eq:30}) one has after
partial integration
\begin{eqnarray}
\nonumber
  <\bar U_0, L_0 U> &=&\lim_{l\rightarrow\infty}{1\over l^2}\int_{-l/2}^{l/2}dx \int_{-l/2}^{l/2}dy
  \biggr[ \int_{-\infty}^{-d}\!\!dz\, \Delta\bar\phi_0^{(3)\ast}\phi^{(3)}
  +{1+\eta\over 1-\eta} \int_{-d}^{0}\!\!dz\, \Delta\bar\phi_0^{(2)\ast}\phi^{(2)}\\
\nonumber
  &&
  +\int_{0}^{\infty}\!\!dz\, \Delta\bar\phi_0^{(1)\ast}\phi^{(1)}
  -\bar\phi_0^{(3)\ast}\bigr|_{-\infty}\partial_z\phi^{(3)}\bigr|_{-\infty}
  +\left(\bar\phi_0^{(3)\ast} -\bar\phi_0^{(2)\ast}\right)\Bigr|_{-d}\partial_z\phi^{(3)}\bigr|_{-d}\\
  \label{eq:64}
  &&
  -\left(\partial_z\bar\phi_0^{(3)\ast} -{1+\eta\over 1-\eta}\partial_z\bar\phi_0^{(2)\ast}\right)
  \biggr|_{-d}\phi^{(3)}\bigr|_{-d}+\bar\phi_0^{(1)\ast}\bigr|_{\infty}\partial_z\phi^{(1)}\bigr|_{\infty}+R\biggr]\; .
\end{eqnarray}
From Eq.~(\ref{eq:64}) it is concluded that the linear adjoint system is governed by
three differential equations of second order for the adjoint potentials $\bar\phi_0^{(i)}$.
According to Eq.~(\ref{eq:64}) the four conditions to be fulfilled by these adjoint potentials
are
\begin{eqnarray}
  \label{eq:65}
  {\rm at~}z=-\infty\qquad
   0 &=& \bar\phi_0^{(3)}\; ,\\
  \label{eq:66}
  {\rm at~}z=-d\qquad\hskip 0.17cm
  0 &=& -\bar\phi_0^{(2)} + \bar\phi_0^{(3)}\; ,\\
  \label{eq:67}
  {\rm at~}z=-d\qquad\hskip 0.17cm
  0 &=& -{1+\eta\over 1-\eta}\partial_z\bar\phi_0^{(2)} + \partial_z\bar\phi_0^{(3)}\; ,\\
  \label{eq:68}
  {\rm at~}z=\infty\qquad\hskip 0.26cm
   0 &=& \bar\phi_0^{(1)}\; ,
\end{eqnarray}
since only terms on the surface should remain. The residual terms
\begin{eqnarray}
\nonumber
R &=& \partial_z\phi^{(2)}\bigr|_0\left( {1+\eta\over 1-\eta}\bar\phi_0^{(2)\ast}+{1+\eta\over 1-
      \eta}\, e\bar u_{0,5}^\ast \right) \biggr|_0
      +\phi^{(2)}\bigr|_0\left( -{1+\eta\over 1-\eta}\partial_z\bar\phi_0^{(2)\ast}
      - d\bar u_{0,4}^\ast\right) \biggr|_0\\
\nonumber
  && +\partial_z\phi^{(1)}\bigr|_0\left[ -\bar\phi_0^{(1)\ast}-e\bar u_{0,5}^\ast 
     -{2(1-\eta)\over \eta(1+\eta)}B_c^2\, f\bar u_{0,6}^\ast \right]\biggr|_0
     +\phi^{(1)}\bigr|_0\left( \partial_z\bar\phi_0^{(1)\ast}+d\bar u_{0,4}^\ast \right) \Bigr|_0\\
\label{eq:69}
  && +\zeta \left[ {2\eta\over 1-\eta}\,d\bar u_{0,4}^\ast+(\partial_{xx}+\partial_{yy}-1)\, f\bar u_{0,6}^\ast
      \right]\biggr|_0
\end{eqnarray}
are bounded to contain only the two remaining boundary conditions and one further term,
all to be determined at $z=0$, for a proper definition of $L_0^\dagger$.
Comparing the terms in $R$ and the surface components of $U$ $\bigr($see Eq.~(\ref{eq:44})$\bigr)$,
one realizes that the first and third term in Eq.~(\ref{eq:69}) have to vanish. As a consequence
of the immediate choice $e\bar u_{0,5} = -\bar\phi_0^{(2)}$, the sixth component
of the adjoint vector $\bar U_0$, $\bar u_{0,6}$, is a linear combination of $\bar\phi_0^{(1)}$
and $\bar\phi_0^{(2)}$, i.e. a linear combination of $\bar\phi_0^{(1)}$ and $\bar u_{0,5}$.
Certainly a result contrary to the condition that the set of variables in $\bar U_0$ have
to be linearly independent. Testing the choice
\begin{equation}
\label{eq:70}
d\bar u_{0,4} = k\bar\phi_0^{(1)}\qquad
e\bar u_{0,5} = -\bar\phi_0^{(2)}\qquad
f\bar u_{0,6} ={-\eta^2(1+\eta)\over (1-\eta)^2 B_c^2}\bar\zeta_0
\end{equation}
results in
\begin{eqnarray}
\nonumber
 <\bar U_0, L_0 U> &=&\lim_{L\rightarrow\infty}{1\over l^2}\int_{-l/2}^{l/2}dx \int_{-l/2}^{l/2}dy
  \biggr[ \int_{-\infty}^{-d}\!\!dz\, \Delta\bar\phi_0^{(3)\ast}\phi^{(3)}
  +{1+\eta\over 1-\eta} \int_{-d}^{0}\!\!dz\, \Delta\bar\phi_0^{(2)\ast}\phi^{(2)}\\
\nonumber
  &&
  +\int_{0}^{\infty}\!\!dz\, \Delta\bar\phi_0^{(1)\ast}\phi^{(1)}
  -\left(- \partial_z\bar\phi_0^{(1)\ast} + {1+\eta\over 1-\eta}\partial_z\bar\phi_0^{(2)\ast}\right)\biggr|_0
   \,\phi^{(2)}\bigr|_0\\
\nonumber
  &&
  -{\eta^2(1+\eta)\over (1-\eta)^2B_c^2}\left( -{2(1-\eta)B_c^2\over \eta (1+\eta)}
   \partial_z\bar\phi_0^{(1)\ast}+(\partial_{xx}+\partial_{yy}-1)\bar\zeta_0^\ast\right)\biggr|_0\,\zeta\\
\label{eq:71}
  &&
  -\left( -{2\eta\over (1-\eta)}\, \zeta_0^\ast +\bar\phi_0^{(1)\ast}- \bar\phi_0^{(2)\ast}
   \right)\biggr|_0\,\partial_z\phi^{(1)}\bigr|_0\biggr]\; ,
\end{eqnarray}
where $\partial_z\bar\phi_0^{(1)}\bigr|_0=-k\bar\phi_0^{(1)}\bigr|_0$ was used. This identity is
justified because of the Laplace equation for $\bar\phi_0^{(1)}$ $\bigr($see last $z$-integral
in Eq.~(\ref{eq:64})$\bigr)$ and the condition (\ref{eq:68}). After inserting (\ref{eq:70})
into Eq.~(\ref{eq:62}), it becomes clear that Eq.~(\ref{eq:71}) has almost the proper form in order
to define the adjoint operator $L_0^\dagger$ via $<\bar U_0, L_0 U> =<L_0^\dagger\bar U_0, U>$.
The last remaining but unsolved problem is posed by the third
surface term in Eq.~(\ref{eq:71}). Instead of $k\phi^{(1)}|_0$, the expression 
$-\partial_z\phi^{(1)}|_0$ appears. Both terms are only equally in the first order of the
expansion, where $\phi_0^{(1)}\sim\expo^{-kz}$ $\bigr($see Eq.~(\ref{eq:53})$\bigr)$. In higher
orders of the expansion, the functions $\phi_1^{(1)}$ and $\phi_2^{(1)}$ containing wave vectors
which are linear combinations of two and three, respectively, wave vectors of the basic modes.
As a result $z$-dependences will appear as
$\phi_1^{(1)}\sim\expo^{-|{\bf k}_n\pm {\bf k}_m|z}$ and
$\phi_2^{(1)}\sim\expo^{-|{\bf k}_n\pm {\bf k}_m\pm {\bf k}_l|z}$, respectively. The absolute value
of the resulting wave vector is usually not equal to $k$. Thus the task
remains that for a nonlinear analysis a linear adjoint operator should be
definable which is valid for a wider set of function as $\expo^{-kz}$.

The generic problem of a nonzero, deformed surface is illustrated by calculating the scalar
product of the augmented state $U$ vector and its adjoint one $\bar U$
\begin{eqnarray}
\nonumber
<\bar U, U> &=&\varepsilon^2 <\bar U_0 , U_0> +\varepsilon^3\left( <\bar U_0 , U_1>
               +<\bar U_1 , U_0>\right) +O(\varepsilon^4)\; ,\\
\nonumber
&=& \lim_{l\rightarrow\infty}{1\over l^2}\int_{-l/2}^{l/2}dx \int_{-l/2}^{l/2}dy
  \biggr[ \int_{-\infty}^{-d}\!\!dz\, \bar\phi^{(3)\ast}\phi^{(3)}
  +{1+\eta\over 1-\eta} \int_{-d}^{\zeta}\!\!dz\, \bar\phi^{(2)\ast}\phi^{(2)}
  +\int_{\zeta}^{\infty}\!\!dz\, \bar\phi^{(1)\ast}\phi^{(1)}\\
\label{eq:72}
  &&
  +d\bar u_4^\ast\bigr|_\zeta\,\phi^{(1)}\bigr|_\zeta
  +e\bar u_5^\ast\bigr|_\zeta\,\phi^{(2)}\bigr|_\zeta
  +f\bar u_6^\ast\bigr|_\zeta\,\zeta\biggr]\; .
\end{eqnarray}
Since $U$ and $\bar U$ contain the complete information of the disturbed state, the unknown
and deformed surface $\zeta$ appears as a bound at two integrals and at the surface contributions.
Whereas the expansion of the three surface contributions can be performed
accordingly to Eq.~(\ref{eq:33}), the expansion of the two integrals involving the deformed
surface $\zeta$ is accomplished as
\begin{equation}
\label{eq:73}
\int_{-d}^{\zeta}\!\!dz\,\bar\phi^{(2)\ast}\phi^{(2)}
=\varepsilon^2\int_{-d}^0\!\!dz\, \phi_0^{(2)\ast} \phi_0^{(2)}
   +\varepsilon^3\left[ \zeta_0 \bar \phi_0^{(2)\ast}\bigr|_0\phi_0^{(2)}\bigr|_0
   +\int_{-d}^0\!\!dz\, \left( \bar\phi_0^{(2)\ast} \phi_1^{(2)}
   +\bar\phi_1^{(2)\ast} \phi_0^{(2)}\right)\right] +O(\varepsilon^4)\; .
\end{equation}
The analogous expansion can be applied to the second integral involving $\zeta$. Matching
the two lowest powers of $\varepsilon$ in Eq.~(\ref{eq:72}), one gets
\begin{eqnarray}
\nonumber
<\bar U_0, U_0> &=& \lim_{l\rightarrow\infty}{1\over l^2}\int_{-l/2}^{l/2}dx \int_{-l/2}^{l/2}dy
  \biggr[ \int_{-\infty}^{-d}\!\!dz\, \bar\phi_0^{(3)\ast}\phi_0^{(3)}
  +{1+\eta\over 1-\eta} \int_{-d}^{0}\!\!dz\, \bar\phi_0^{(2)\ast}\phi_0^{(2)}
  +\int_{0}^{\infty}\!\!dz\, \bar\phi_0^{(1)\ast}\phi_0^{(1)}\\
\label{eq:75}
  &&
  +d\bar u_{0,4}^\ast\bigr|_0\,\phi_0^{(1)}\bigr|_0
  +e\bar u_{0,5}^\ast\bigr|_0\,\phi_0^{(2)}\bigr|_0
  +f\bar u_{0,6}^\ast\bigr|_0\,\zeta_0\biggr]
\end{eqnarray}
and
\begin{eqnarray}
\nonumber
<\bar U_0, U_1> + <\bar U_1, U_0>&=&
  \lim_{l\rightarrow\infty}{1\over l^2}\int_{-l/2}^{l/2}dx \int_{-l/2}^{l/2}dy
  \biggr\{ \int_{-\infty}^{-d}\!\!dz\, \left(\bar\phi_0^{(3)\ast}\phi_1^{(3)} + \bar\phi_1^{(3)\ast}\phi_0^{(3)}\right)
  \\
&&
\nonumber
  +{1+\eta\over 1-\eta} \int_{-d}^{0}\!\!dz\, \Bigr(\bar\phi_0^{(2)\ast}\phi_1^{(2)}
  +\bar\phi_1^{(2)\ast}\phi_0^{(2)}\Bigr)
  +\int_{0}^{\infty}\!\!dz\, \left(\bar\phi_0^{(1)\ast}\phi_1^{(1)}+ \bar\phi_1^{(1)\ast}\phi_0^{(1)}\right)\\
&&
\nonumber
  +d\left(\bar u_{0,4}^\ast\,\phi_1^{(1)}+\bar u_{1,4}^\ast\,\phi_0^{(1)}\right)\Bigr|_0
  +e\left(\bar u_{0,5}^\ast\,\phi_1^{(2)}+\bar u_{1,5}^\ast\,\phi_0^{(2)}\right)\Bigr|_0\\
&&
\nonumber
  +f\left(\bar u_{0,6}^\ast\bigr|_0\,\zeta_1 +\bar u_{1,6}^\ast\bigr|_0\,\zeta_0\right)
  +\zeta_0\Bigr[\bar\phi_0^{(2)\ast}\phi_0^{(2)} -\bar \phi_0^{(1)\ast}\phi_0^{(1)}\\
\nonumber
&&
  +d\left( \bar u_{0,4}^\ast\,\partial_z\phi_0^{(1)} +\partial_z\bar u_{0,4}^\ast\,\phi_0^{(1)}\right)
  +e\left( \bar u_{0,5}^\ast\,\partial_z\phi_0^{(2)} +\partial_z\bar u_{0,5}^\ast\,\phi_0^{(2)}
  \right)\Bigr]\Bigr|_0\\
\label{eq:76}
&&
  +f\partial_z\bar u_{0,6}^\ast\bigr|_0\,\zeta_0^2\biggr\}\; .
\end{eqnarray}
As discussed for Eq.~(\ref{eq:71}), the problems are appearing beyond the first order of expansion.
The form for the scalar product of $<\bar U_0, U_0>$ is the expected one
$\bigr($Eq.~(\ref{eq:75})$\bigr)$. In the second order expansion, the left hand side of
Eq.~(\ref{eq:76}) lets await only products of functions which belong to the first and
second order of expansion. But the expansion on the deformed surface
generates additional terms, where each one is formed by three functions of the
first order expansion. These additional terms can neither be assigned to $<\bar U_0, U_1>$
nor to $<\bar U_1, U_0>$ by simple arguments. Similar dilemmas are present for the third
and any higher expansion order with many more not-assignable terms. Neglecting surface deformations
is certainly not an option for the solution of this problem.

It becomes now apparent why the derivation of an amplitude equation in
\cite{bragard98,engel00_mara,deyirmenjian97} with the approximation of a {\it plane} surface
succeeded. The first order expansion on a {\it deformed} surface is equivalent to that
approximation of a plane surface. Therefore the problems of not-assignable terms stemming from
higher orders in the expansion do not occur.

\section{Discussion}
\label{sec:5}
The detailed analysis of the last section showed that the definition of a 
linear adjoint operator for pure surface-nonlinearities has not yet been
accomplished. Two attempts which led to unsatisfactory consequences
were described here. They point to the fact that in the known literature
no derivation can be traced in which an expansion on the deformed surface was
taken into account {\it and} a linear adjoint operator was defined. The latter is necessary
to determine higher order terms of the external parameter; for the system considered 
here these are $B_1$ and $B_2$, see Eq.~(\ref{eq:33}). This fact is surprisingly since pattern
forming systems with a deformed surface belong to the set of classical hydrodynamical systems.

For the nonlinear analysis of instabilities in MF, a very similar expansion route was followed
in \cite{malik84,malik93,khosla98}. Remarkably, only very shortly the existence of an adjoint problem is
mentioned in \cite{malik93,khosla98}. No details with respect to the definition of a scalar
product and a linear adjoint operator were published. The same expansion as in Eq.~(\ref{eq:32})
is used, but the external parameter, the magnetic field $H$, is not expanded accordingly to
Eq.~(\ref{eq:31}). The latter step is a inconsistency since the multiple-scale expansion comes along
with the expansion of the physical quantities {\it and} the external driving parameters as the
Rayleigh number \cite{busse82,godreche98,zebib96} or the Marangoni number
\cite{golovin97,bragard98,engel00_mara}.
The gain of the lacking expansion is that higher order terms of the external parameter
have not to be determined. The solvability condition in second order 
postulated  in \cite{malik84,malik93} demands that the amplitude has to have a nonvanishing
derivative everywhere with respect to the slow spatial variable. That means that the amplitude
is a strictly monotonously increasing or decreasing function which is a very special type of solution.
Selection problems between regular patterns can not be tackled with such a type of solution.

The problems caused by a deformed surface for the derivation of an amplitude equation are not
unique to instabilities of magnetic fluids. The analysis
of the B\' enard-Marangoni convection or of parametric surface (Faraday) waves is confronted
with similar difficulties. For the B\' enard-Marangoni convection the surface deformation is often
disregarded in fluid-gas systems \cite{bragard98,rosenblat82} and two-fluid systems
\cite{engel00_mara}. In \cite{kraska79,davis87} a surface deflection was imitated by a nonzero
Crispation number, thus avoiding the expansion of any quantity on a deformed surface.
In \cite{golovin97} the expansion of the layer thickness was tuned in
such a way that eigenfunctions could be used which correspond to the case of an undeformed
interface.

In the nonlinear analysis of Faraday waves an explicit evaluation of quantities on
the free surface is rare. In \cite{chen99} the driving parameter is not expanded, similar
to \cite{malik84,malik93,khosla98}. No expansion on the surface is performed, but
two different amplitudes are introduced for the solution of first and third order.
The adjoint operator is circumvented in this way. The two different amplitudes are
just the two quantities needed to attach use to the solvability conditions. 
Under the constrains of an inviscid fluid and
an irrotational flow field, Milner \cite{milner91} derived an amplitude equation for parametrically
driven capillary waves. An expansion of the velocity potential on the surface was included, but an
adjoint problem was not formulated.

The overall picture is that several circumventing routes were used in order to avoid the
explicit evaluation of quantities on a deformed surface and the formulation of a linear
adjoint operator. The used alternatives were hardly motivated which is why the real reasons
behind the search for alternatives are not publicly known. Instead of using the
solvability conditions to determine higher order terms of the external parameters, they were
converted to constraining conditions for the amplitudes of the expanded physical quantities.
This presents a rather unsatisfactory situation.

The deeper reasons for this unsatisfactory situation are disclosed in this article.
The adjoint problem  is presented in detail for nonlinearities purely triggered by a deformed
surface as in the case of the Rosensweig instability. By expanding the physical quantities and the
external parameter, the knowledge of the linear adjoint operator is essentially to proceed
towards the nonlinear amplitude equation. The main and still unsolved problem is the proper
definition of the linear adjoint operator. Since this problem mounts a principal
barrier and is not mentioned as a reason for the search of alternative approaches, it
is presented here even if no solution can presently be offered. This situation
entails one foremost question: Is it possible to apply the concept
of Fredholms theorem to pure surface-nonlinearities? If yes, what is the adapted route for
defining a linear adjoint operator. If not, what is a generic and mathematically proven alternative
since the hitherto used routes are lacking these features. Therefore further analyses need to be
done in order to solve this problem and to clarify the open questions.

If these problems for the Rosensweig instability, caused by the single external
excitation of a magnetic field, are solved, other instabilities caused by
different external excitations can be fruitful tackled. Phenomena as
standing twin peaks \cite{reimann99} or domain structures \cite{mahr98_para}
for parametrically excited MF under the influence of a magnetic field are
designated future examples for an analysis by amplitude equations derived from the
basic hydrodynamic equations.

\section*{Acknowledgments}
\label{sec:6}
The author has much benefited from discussion with A. Engel and R. Friedrichs and like to
thanks F. H. Busse, H. W. M\"uller, M. Petry, and J. Weilepp for stimulating correspondence.
This work was supported by the Deutsche Forschungsgemeinschaft under Grant No.
LA 1182/2 and EN 278/2.

\bibliographystyle{prsty}
\bibliography{mf_viscous,mf_general,mf_AE,mf_conv,marangoni,electroconvection,AE_general,conv_shell,faraday}

\begin{thebibliography}{10}

\bibitem{odenbach00_viscous}
S. Odenbach, Int. J. Mod. Phys. B {\bf 14},  1615  (2000).

\bibitem{rosensweig}
R.~E. Rosensweig, {\em Ferrohydrodynamics} (Cambridge University Press,
  Cambridge, 1985).

\bibitem{cowley67}
M.~D. Cowley and R.~E. Rosensweig, J. Fluid Mech. {\bf 30},  671  (1967).

\bibitem{salin93}
D. Salin, Europhys. Lett. {\bf 21},  667  (1993).

\bibitem{weilepp96}
J. Weilepp and H.~R. Brand, J. Phys. II France {\bf 6},  419  (1996).

\bibitem{abou00}
B. Abou, J. Wesfreid, and S. Roux, J. Fluid Mech. {\bf 416},  217  (2000).

\bibitem{lange00_wave}
A. Lange, B. Reimann, and R. Richter, Phys. Rev. E {\bf 61},  5528  (2000).

\bibitem{petrelis00}
F. P{\' e}tr{\' e}lis, {\' E}. Falcon, and S. Fauve, Eur. Phys. J. B {\bf 15},
  3  (2000).

\bibitem{browaeys01}
J. Browaeys, J. Bacri, R. Perzynski, and M.~I. Shliomis, Europhys. Lett. {\bf
  53},  209  (2001).

\bibitem{kuznetsov76}
E.~A. Kuznetzov and M.~D. Spektor, Sov. Phys. JETP {\bf 44},  136  (1976).

\bibitem{gailitis77}
A. Gailitis, J. Fluid Mech. {\bf 82},  401  (1977).

\bibitem{beyer80}
K. Beyer, ZAMM {\bf 60},  235  (1980).

\bibitem{twombly80}
E.~E. Twombly and J.~W. Thomas, IEEE Trans. Magn. {\bf 16},  214  (1980).

\bibitem{twombly83}
E.~E. Twombly and J.~W. Thomas, SIAM J. Math. Anal. {\bf 14},  736  (1983).

\bibitem{silber88}
M. Silber and E. Knobloch, Physica D {\bf 30},  83  (1988).

\bibitem{friedrichs01}
R. Friedrichs and A. Engel, Phys. Rev. E {\bf 64},  021406  (2001).

\bibitem{herrero94}
H. Herrero, C. P{\' e}rez-Garc{\'\i}a, and M. Bestehorn, Chaos {\bf 4},  15
  (1994).

\bibitem{kubstrup96}
C. Kubstrup, H. Herrero, and C. P{\' e}rez-Garc{\'\i}a, Phys. Rev. E {\bf 54},
  1560  (1996).

\bibitem{malik84}
S.~K. Malik and M. Singh, Quart. Appl. Math. {\bf XLII},  359  (1984), there is
  a misprint in equation (9).

\bibitem{malik93}
S.~K. Malik and M. Singh, Quart. Appl. Math. {\bf LI},  519  (1993).

\bibitem{engel99_non}
A. Engel, H. Langer, and V. Chetverikov, J. Magn. Magn. Mat. {\bf 195},  212
  (1999).

\bibitem{busse82}
F.~H. Busse and N. Riahi, J. Fluid Mech. {\bf 123},  283  (1982).

\bibitem{godreche98}
{\em Hydrodynamics and nonlinear instabilities}, edited by C. Godr{\' e}che and
  P. Manneville (Cambridge Univ. Press, Cambridge, 1998), sec. 4.

\bibitem{zebib96}
A. Zebib, J. Fluid Mech. {\bf 321},  121  (1996).

\bibitem{golovin97}
A.~A. Golovin, A.~A. Nepomnyashchy, and L.~M. Pismen, J. Fluid Mech. {\bf 341},
   317  (1997).

\bibitem{bragard98}
J. Bragard and M.~G. Velarde, J. Fluid Mech. {\bf 368},  165  (1998).

\bibitem{engel00_mara}
A. Engel and J.~B. Swift, Phys. Rev. E {\bf 62},  6540  (2000).

\bibitem{deyirmenjian97}
V.~B. Deyirmenjian, Z.~D. Daya, and S.~W. Morris, Phys. Rev. E {\bf 56},  1706
  (1997).

\bibitem{abou97}
B. Abou, G.~N. de~Surgy, and J.~E. Wesfreid, J. Phys. II France {\bf 7},  1159
  (1997).

\bibitem{mueller98}
H.~W. M\"uller, Phys. Rev. E {\bf 58},  6199  (1998).

\bibitem{lange01_growth}
A. Lange, Europhys. Lett. {\bf 55},  327  (2001).

\bibitem{cross93}
M.~C. Cross and P.~C. Hohenberg, Rev. Mod. Phys. {\bf 65},  851  (1993).

\bibitem{khosla98}
H.~K. Khosla and S.~K. Malik, Phys. Fluids {\bf 10},  1962  (1998).

\bibitem{rosenblat82}
S. Rosenblat, S.~H. Davis, and G.~H. Homsy, J. Fluid Mech. {\bf 120},  91
  (1982).

\bibitem{kraska79}
J.~R. Kraska and R.~L. Sani, Int. J. Heat Mass Transfer {\bf 22},  535  (1979).

\bibitem{davis87}
S.~H. Davis, Ann. Rev. Fluid Mech. {\bf 19},  403  (1987).

\bibitem{chen99}
P. Chen and J. Vi{\~ n}als, Phys. Rev. E {\bf 60},  559  (1999).

\bibitem{milner91}
S.~T. Milner, J. Fluid Mech. {\bf 225},  81  (1991).

\bibitem{reimann99}
B. Reimann, T. Mahr, R. Richter, and I. Rehberg, J. Magn. Magn. Mat. {\bf 201},
   303  (1999).

\bibitem{mahr98_para}
T. Mahr and I. Rehberg, Phys. Rev. Lett. {\bf 81},  89  (1998).

\end{thebibliography}

\end{document}